# Objects in JWST's mirrors are closer than they appear

**Stephen Serjeant[1], Tom J.L.C. Bakx[2]**
[1] School of Physical Sciences, The Open University, Milton Keynes, MK7 6AA, UK
[2] Department of Space, Earth, & Environment, Chalmers University of Technology, Chalmersplatsen 4 412 96 Gothenburg, Sweden

**The James Webb Space Telescope (JWST) has revealed extremely distant galaxies at unprecedentedly early cosmic epochs from its deep imaging using the technique of photometric redshift estimation (e.g., [1]), with its subsequent spectroscopy confirming their redshifts unambiguously (e.g., [2]), demonstrating the ability of JWST to probe the earliest galaxies, one of its major scientific goals. However, as larger samples continue to be followed up spectroscopically, it has become apparent that nearly all photometric redshifts at these epochs are biased high with confidence >>99%, for as yet unclear reasons. Here we show that this is the same statistical effect that was predicted in different contexts by Sir Arthur Eddington[3] in 1913, in that there exist more lower redshift galaxies to be scattered upwards than the reverse. The bias depends on the shape of the intrinsic redshift distribution, but as an approximate heuristic, all ultra-high photometric redshift estimates must be corrected downwards by up to one standard deviation.**

In this work we use a recent spectroscopic redshift compilation[4] comprising 26 galaxies with redshifts of $z_{spec}$=8.61 to 13.20. This is the largest and most comprehensive ultra-high-redshift spectroscopic sample to date and includes the spectroscopic confirmations of the four galaxies in Ref.[1]. The galaxies in this compilation were initially selected via the inferred presence of the Lyman absorption break, in which the intervening neutral intergalactic medium absorbs essentially all photons at energies above the $n$=2-1 atomic hydrogen transition. These redshifts are well within the reionization epoch, before population III stars, OB stars and active nuclei reionized their surroundings, so the overwhelming preponderance of neutral gas is predicted to generate a very strong Lyman decrement. We include the galaxy GN-z11, with a photometric redshift estimate[6] of $z_{phot} = 11.09^{+0.08}_{-0.12}$ from grism spectroscopy binned to a coarse wavelength resolution similar to those of Hubble Space Telescope (HST) medium-width passbands. We exclude one galaxy from the compilation, GS+53.11243-27.77461, because it was discovered in blind spectroscopy rather than being pre-selected photometrically, but add the spectroscopic confirmation[5] of the galaxy GLASS-z12.

In Figure 1, we compare the spectroscopic redshifts with the original photometrically-estimated redshifts[4]. The photometric redshifts are subtly but strikingly systematically overestimating the accurate spectroscopic measurements. The photometric redshift uncertainties are often asymmetric; in about half of our sample the lower error is worse, while in the rest of the sample the upper error is equal or worse. Of the 26 galaxies in this sample, 21 show photometric redshift overprediction. We can reject the null hypothesis of a *p*=0.5 binomial distribution at 99.8% confidence. The mean of the histogram in Figure 1 is offset from zero with a significance 4.1σ, equivalent to 99.9998% confidence. An Anderson-Darling test rejects a Gaussian with zero mean and unit variance at a significance level of ~2×10$^{-8}$.

Eddington[3] argued for a systematic bias in the number counts of stars arising from the effect of measurement uncertainties. This bias is well-known to affect faint submillimetre galaxy surveys. The same principle applies here. The luminosity function evolves steeply with redshift at these epochs, not only with a (1+z)$^4$ decline in luminosity density and comoving volume elements *dV/dz* also scaling inversely approximately as (1+z), but also with a steep bright-end luminosity function slope[7], and with luminosities at a given flux density also scaling approximately as (1+z)$^{1.5}$ it is likely that each ultra-high-redshift galaxy is sampled from a population with extremely steep intrinsic number counts. The observed compilation in Ref.[4], however, is a heterogeneous compilation with a diverse range of selection effects between objects, so the observed N(z) in Ref.[4] need not reflect the steepness of the underlying number counts or luminosity function in an obvious way; indeed, the largest predictions at *z*~13 and *z*~15 for each survey in Ref.[8] imply intrinsic $dN/dz \propto (1+z)^{-k}$ with $k = 19 \pm 6$.

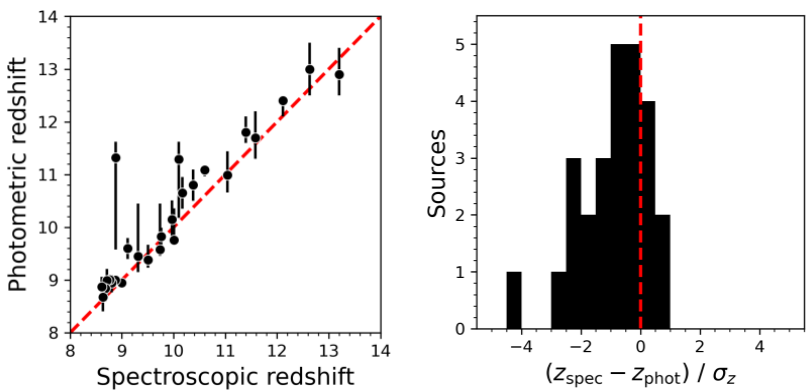

*Figure 1: The left panel compares the photometric redshift estimates against the precise spectroscopic measurements in our ultra high-redshift sample, including the four galaxies in Ref.[21]. The red dashed line shows the one-to-one relation; note that the data trends above this line. The right panel shows the histogram of photometric redshift overestimation divided by the uncertainties in photometric redshift, with the red dashed line again indicating no bias; note the striking offset to the left.*

Each ultra-high-redshift galaxy will therefore be selected from an intrinsic number count *dN/dz* that has been convolved with the photometric redshift probability distribution function. An exact analytic form does not exist, but we find the mean offset of photometric and exact spectroscopic redshifts can be fairly well approximated by

$$\langle z_{spec} - z_{phot} \rangle \approx -0.949 \, k^{1.03} z_{phot}^{-1.01} \sigma^{2.03} \qquad (1)$$

where the intrinsic *dN/dz* is parameterised as being a $(1+z)^{-k}$ power law, and σ is a symmetrical uncertainty in photometric redshift. This analytic prediction agrees to within 20% of the full numerical calculation for $7.5 \leq z_{phot} \leq 26.5$, $0.1 \leq \sigma \leq 0.75$ and $5 \leq k \leq 24$. This approximate relation is roughly linear in *k* but has a surprisingly strong dependence on σ. A Schechter luminosity function would further curve the redshift distribution downwards at the highest redshifts, making the Eddington bias stronger, similarly to how magnification bias preferentially affects the number counts of the brightest submillimetre galaxies.

Furthermore, the photometric redshift uncertainties are themselves likely to be understated due to limitations in the templates used. The presence or absence of the Lyα emission line results in a mean shift of 0.07 in photometric redshift[9], while the discovery of a Lyα absorption damping wing (e.g. Ref.[10]) can shift the estimates by 0.06 in the opposite direction. At lower redshifts, template-based systematics can easily approach[11] 0.02(1+z). Depending on how likelihoods or posteriors are implemented, there may also be stochastic biases[13] at a similar level to the random errors driven by the asymmetric uncertainties. In partial support of our argument that photometric redshift estimates tend to subtly understate their uncertainties, the JWST spectrum[12] of Lyα in GN-z11 is curiously inconsistent with the apparent position of the spectral break in the earlier HST grism data[6], suggesting either underestimated uncertainties or systematics in the difficult subtraction of foreground contaminants in the HST grism data, and/or insufficient signal-to-noise per spaxel to detect the extended Lyα, and/or insufficient diversity in templates to capture the possible spatio-spectral variation in continuum, HI opacity and Lyα emission.

As an illustrative exercise, we find that the distribution of $(z_{spec}-z_{phot})/\sigma$ can be made consistent with a Gaussian distribution with zero mean and unit variance if the $z_{phot}$ values are corrected assuming k=19 (consistent with Ref.[8]) and a template-based additional systematic uncertainty[9,10] of σ(z)=±0.1. We conclude that Eddington bias, together with a subtle underestimate in photometric redshift uncertainties, can entirely account for the systematic shift. Our recommendation is that future photometric redshift works incorporate a model intrinsic dN/dz into Bayesian priors, with both biases and uncertainties tested against simulations.

The authors declare no competing interests.

**References**
1. Robertson, B.E., et al., Identification and properties of intense star-forming galaxies at redshifts z > 10. *Nature Astronomy*, **7**, 611–621 (2023)